\documentclass[letter]{aa}  

\usepackage{graphicx}
\usepackage{txfonts}
\usepackage{color}
\usepackage{soul}
\begin{document}

\newcommand{\msun}{\rm M_\odot}
\newcommand{\kms}{\rm km\, s^{-1}}
\newcommand{\dd}{\rm d}

\title{On the low ortho-to-para H$_2$ ratio in star-forming filaments}
   \author{
   Alessandro Lupi\inst{1,2}\fnmsep\thanks{alessandro.lupi@unimib.it}
   \and 
   Stefano Bovino\inst{3}
   \and
   Tommaso Grassi\inst{4}
   }
    \institute{
    Dipartimento di Fisica ``G. Occhialini'', Universit\`a degli Studi di Milano-Bicocca, Piazza della Scienza 3, I-20126 Milano, Italy
    \and
    INFN – Sezione di Milano-Bicocca, Piazza della Scienza 3, I-20126 Milano, Italy
    \and 
    Departamento de Astronom\'ia, Facultad Ciencias F\'isicas y Matem\'aticas, Universidad de Concepci\'on, Av. Esteban Iturra s/n Barrio Universitario, Casilla 160, Concepci\'on, Chile
    \and
    Center for Astrochemical Studies, Max-Planck-Institut f\"ur Extraterrestrische Physik,  Giessenbachstrasse 1, D-85748, Garching, Germany
    }

   \date{Received XXX; accepted \today}

  \abstract
  {The formation of stars and planetary systems is a complex phenomenon, which relies on the interplay of multiple physical processes. Nonetheless, it represents a crucial stage for our understanding of the Universe, and in particular of the conditions leading to the formation of key molecules (e.g. water) on comets and planets. \textit{Herschel} observations demonstrated that stars form out of gaseous filamentary structures in which the main constituent is  
    molecular hydrogen (H$_2$). Depending on its nuclear spin H$_2$ can be found in two forms: `ortho' with parallel spins and `para' where the spins are anti-parallel. The relative ratio among these isomers, i.e.\,the ortho-to-para ratio (OPR), plays a crucial role in a variety of processes related to the thermodynamics of star-forming gas and to the fundamental chemistry affecting the deuteration of water in molecular clouds, commonly used to determine the origin of water in Solar System's bodies. 
   Here, for the first time, we assess the evolution of the OPR starting from the warm neutral medium, by means of state-of-the-art three-dimensional  magneto-hydrodynamic simulations of turbulent molecular clouds. Our results show that
    star-forming clouds exhibit a low OPR ($\ll 0.1$) already at moderate densities ($\sim$1000 cm$^{-3}$). We also constrain the cosmic rays ionisation rate,  finding that $10^{-16}$\, s$^{-1}$ is the lower limit required to explain the observations of diffuse clouds. Our results represent a step forward in the understanding of the star and planet formation process providing a robust determination of the chemical initial conditions for both theoretical and observational studies.}


   \maketitle
%
%-------------------------------------------------------------------

\section{Introduction}
Star formation is one of the greatest open problems in astrophysics  \citep{Bergin2007,SF2007}, and despite the huge progress made over the last decades, both observationally and theoretically, some fundamental questions still remain open.
Star formation occurs within molecular clouds (MCs), dense and cold regions within galaxies mainly composed of molecular hydrogen (H$_2$). Within these clouds, stars form out of gaseous filamentary structures \citep{Andre2010,Molinari2010} which show quasi-universal average properties \citep{Arzoumanian2011}. 
Due to the huge dynamic range involved and the variety of complex physical processes that couple together on different scales, studying star formation from ab-initio conditions is still unfeasible. For this reason, the problem has been tackled from different sides, i.e.\,on MC scales, in the aim at describing the formation of filaments and cores \citep[e.g.][]{Federrath10,Padoan11,Federrath16b,Padoan2016}, or on the small scales typical of these substructures, neglecting the large scale environment and focussing on the last stages of the gravitational collapse \citep[e.g.][]{Bovino2019,Bovino2020a}. 
Unfortunately, while small scale simulations are extremely useful to study the final stages of the gravitational collapse and compare the results with observations of protostellar cores, the detailed chemical conditions at the onset of gravitational collapse are still very uncertain, and can be constrained only via self-consistent studies on MC scales.

From a chemical point of view, the evolution of species like CO and H$_2$ up to the formation of filaments is crucial to assess the formation  of key molecules (e.g.\,water) at later stages as on comets and planets \citep{WaterSci2011,Bergin2012,Ceccarelli2014,Jorgensen2020}. This process strongly depends on the conditions of the main constituent of star-forming filaments, i.e.\,molecular hydrogen (H$_2$), which can be found in two forms: \emph{ortho} and \emph{para} states, where the spins are parallel or anti-parallel respectively. 
The relative abundance of these isomers has profound implications for both thermodynamic and chemical processes that affect the formation and deuteration of water \citep{Furuya2015,Jensen2021} and its inheritance in the Solar system \citep{Altwegg2014}.

In this work, we assess for the first time the evolution of the OPR starting from large-scale conditions, i.e.\,the warm neutral medium, by means of state-of-the-art three-dimensional (3D) magneto-hydrodynamic (MHD) simulations of turbulent molecular clouds. 
The paper is organised as follows: in Section~\ref{sec:setup} we introduce the setup of our simulations, in Section~\ref{sec:results} we discuss our results, and in Section~\ref{sec:conclusions} we draw our conclusions.

\section{Numerical setup}
\label{sec:setup}
The simulations presented in this work have been performed with the publicly available MHD code \textsc{gizmo} \citep{Hopkins2015,Hopkins2016a,Hopkins2016b}, descendant of \textsc{gadget2} \citep{Springel2005}, which included the gas self-gravity.\footnote{In this study, we employ  a cubic spline kernel with an effective number of neighbours of~32.}
\subsection{Microphysics}
For the purpose of this study, we equipped the code with an on-the-fly non-equilibrium chemistry network, implemented via the public chemistry library \textsc{krome} \citep{Grassi2014}. 
The chemical network we employ is based on \citet{Grassi2017}, which is an updated version of that in \citet{Glover2010}. We include isomer-dependent chemistry, by employing the most up-to-date reaction rates \citep{Sipila2015,Bovino2019}. The final network includes 40 species: H, H$^+$, He, He$^+$, He$^{++}$, ortho-H$_2$, para-H$_2$, ortho-H$_2^+$, para-H$_2^+$, H$^-$, C$^+$, C, O$^+$, O, OH, HOC$^+$, HCO$^+$, CO, CH, CH$_2$, C$_2$, HCO, H$_2$O, O$_2$, ortho-H$_3^+$, para-H$_3^+$, CH$^+$, CH$_2^+$ , CO$^+$, CH$_3^+$ , OH$^+$, H$_2$O$^+$, H$_3$O$^+$, O$_2^+$ , C$^-$, O$^-$, electrons, plus GRAIN0, GRAIN-, and GRAIN+, which represent dust grains. 
A total of 397~reactions connects all these species, including ortho-to-para conversion by protons collisions (H$^+$ and H$_3^+$), adsorption and desorption of CO and water on the surface of grains \citep{Cazaux2010,Hocuk14}, ionisation/dissociation induced by impact with cosmic-rays, and dissociation of molecules induced by a standard interstellar radiation field (Draine flux, \citealt{Draine1978}), which includes self-shielding of H$_2$ \citep{Glover2010} and CO \citep{Visser2009}. Electron attachment and recombination of positive ions on grains are also included in the chemical network \citep{Walmsley2004}, with the Coulomb factor consistently calculated for the Draine flux \citep{Draine1987}, as well as H$_2$ formation on dust \citep[assuming an initial OPR of 3 at formation; see][]{Watanabe2010,Gavilan2012,Hama2013,Wakelam2017} and dust cooling \citep{Grassi2014}, which are determined via pre-computed dust tables, in this case density-, temperature-, and $A_v$-dependent. The extinction parameter is defined as $A_{\rm v} = (10^{-3}n_{\mathrm{H}_2})^{\alpha}$, with $\alpha = 2/3$ \citep[see][]{Grassi2014}, and the H$_2$ column density entering the H$_2$ self-shielding factor as $N_{\rm H_2}=1.87\times 10^{21} A_{\rm v}\,\rm cm^{-2}$ \citep{Grassi2014}.

To consistently follow the thermodynamics of the gas, we further include: metal line cooling from CI, CII, and OI, CO rotational cooling, chemical cooling induced by endothermic reactions, H$_2$ roto-vibrational cooling, Compton cooling, continuum, plus chemical heating induced by exothermic reactions, photoheating, and cosmic rays-induced heating. Photoelectric heating is also included \citep{Bakes1994}, following recent modifications \citep{Wolfire2003}. A floor of 10\,K is imposed. In order to model cosmic-ray attenuation through the cloud, in our fiducial model we employ a variable cosmic-ray flux which depends on local column density, as in \citet{Padovani2018} (see Appendix~\ref{app:methods} for details).

\subsection{Initial conditions}
The region we simulate is a cubic box of 200\,pc filled with homogeneous atomic gas and dust at a hydrogen nuclei density $n_{\rm H,tot}=5\rm\, cm^{-3}$, that corresponds to a total mass of $1.25\times 10^6\rm\, \msun$, i.e.\,a typical giant molecular cloud. The mass and spatial resolution are $0.2\rm\,\msun$ and $\sim 60$~AU respectively, which allow us to properly resolve the formation of observed filaments and clumps with typical masses of $100-1000\rm\,\msun$ and sizes of a few parsecs. Our simulations evolve the gas according to the ideal MHD equations, starting from an initially constant magnetic field of $3\mu$G aligned with the $x$ direction. After an initial relaxation phase aimed at reaching a steady-state turbulence, we turn on self-gravity and on-the-fly non-equilibrium chemistry, the latter including ortho- and para- forms of H$_2$, gas-grain interactions, photochemistry, and cosmic-ray induced reactions (see Appendix~\ref{app:methods} for details). 

\subsection{Filament identification and analysis}
During the evolution, filaments continuously form and disperse, up to the point at which gravity overcomes the thermal, turbulent, and magnetic support, resulting in the decoupling of the structure from the entire cloud, and the beginning of the collapse phase. In order to infer their properties over time, we identify them from two-dimensional (2D) H$_2$ column density maps using \textsc{astrodendro}, imposing a minimum background density $N_{\rm H_2}=10^{21}\rm\, cm^{-2}$ and an rms error of $\sigma_{\rm N}=3\times 10^{20}\rm\, cm^{-2}$. In addition, we require a minimum area of at least 10~pixels. Among the found structures, we assume as filaments only the main branches of the dendrogram, excluding sub-branches and leaves in the hierarchy that likely represent clumps and cores within the filaments. The average properties of the identified structures are then computed by averaging the corresponding 2D maps pixel by pixel. For instance, for H$_3^+$, we employ column density maps integrated along the line-of-sight (the $z$ axis), whereas for magnetic field, temperature, and velocity dispersion, we employ the H$_2$ density-weighted line-of-sight average maps, one per direction depending on the property considered. Notice that computing the OPR using column densities or number densities can results in large differences, due to the dilution effect caused by averaging over different regions along the line-of-sight (as also discussed in the main text). The filament mass is derived from the H$_2$ column density as $M=\sum_i N^i_{\rm H_2}\times \dd S$ with $\dd S$ is the pixel area and the sum is over the pixels associated to the filament. The Mach number $\mathcal{M}\equiv c_s/\sigma_{\rm v}$ is determined from the isothermal sound speed $c_s=\sqrt{k_{\rm B} T_{\rm gas}/(\mu m_{\rm H})}$, with $m_{\rm H}$ the proton mass, $k_{\rm B}$ the Boltzmann constant, and $\mu=2.4$ the molecular weight, and the average 3D velocity dispersion $\sigma_{\rm v}=\sqrt{\sum_{k=x,y,z}\sigma_{{\rm v},k}^2}$, where $\sigma_{{\rm v},k}$ is the average over the filament of the $k-$th direction average velocity dispersion. Since we are not interested into an extremely accurate measure of the filament lengths and widths, and given the complex geometry of our simulated filaments, far from a perfect cylinder, we opt for a simple and approximate determination of the filament length and width. In detail, we proceed as follows: we first align the structure along its major axis (using the position angle of the ellipse associated to the branch of the dendrogram) and then define the length $L$ as the maximum horizontal distance among the pixels belonging to the filament. The width $W$ is then retrieved as $W=A_{\rm px}/L$, which guarantees that the area is preserved exactly, and the error in the estimate of $W$ is as good as that used for $L$. At last, a crucial parameter used to determine the fate of (potentially) star-forming filaments is the mass-to-length ratio $M/L$, which is typically compared to a critical value $[M/L]_{\rm c}=2c_s/{\rm G}$, with $G$ the gravitational constant.

\section{Results}
\label{sec:results}
We evolve the cloud for a few Myr, necessary for the first filaments to form out of the low-density material. During this stage, the temperature evolves self-consistently (see Appendix~\ref{app:relax}), leading to an average Mach number in the cloud of $\sim 5-6$ after 4~Myr, consistent with typically observed values \citep{MacLow04}.
The gas distribution in our fiducial simulated cloud at 4~Myr (just before the formation of the first sink particle, see Appendix~\ref{app:methods}) is shown in Fig.~\ref{fig:Nmap}, with the four panels on the right reporting the line-of-sight-integrated ortho-to-para ratio ${\rm OPR}^N\equiv N_{\rm o-H_2}/N_{\rm p-H_2}$ of four massive filamentary structures out of hundreds identified. Qualitatively, Fig.~\ref{fig:Nmap} indicates that the OPR is around 0.1 already at these early stages, with peaks of $\sim 0.01$ in the densest regions (clumps). 

\begin{figure*}
\centering
\includegraphics[width=0.8\textwidth]{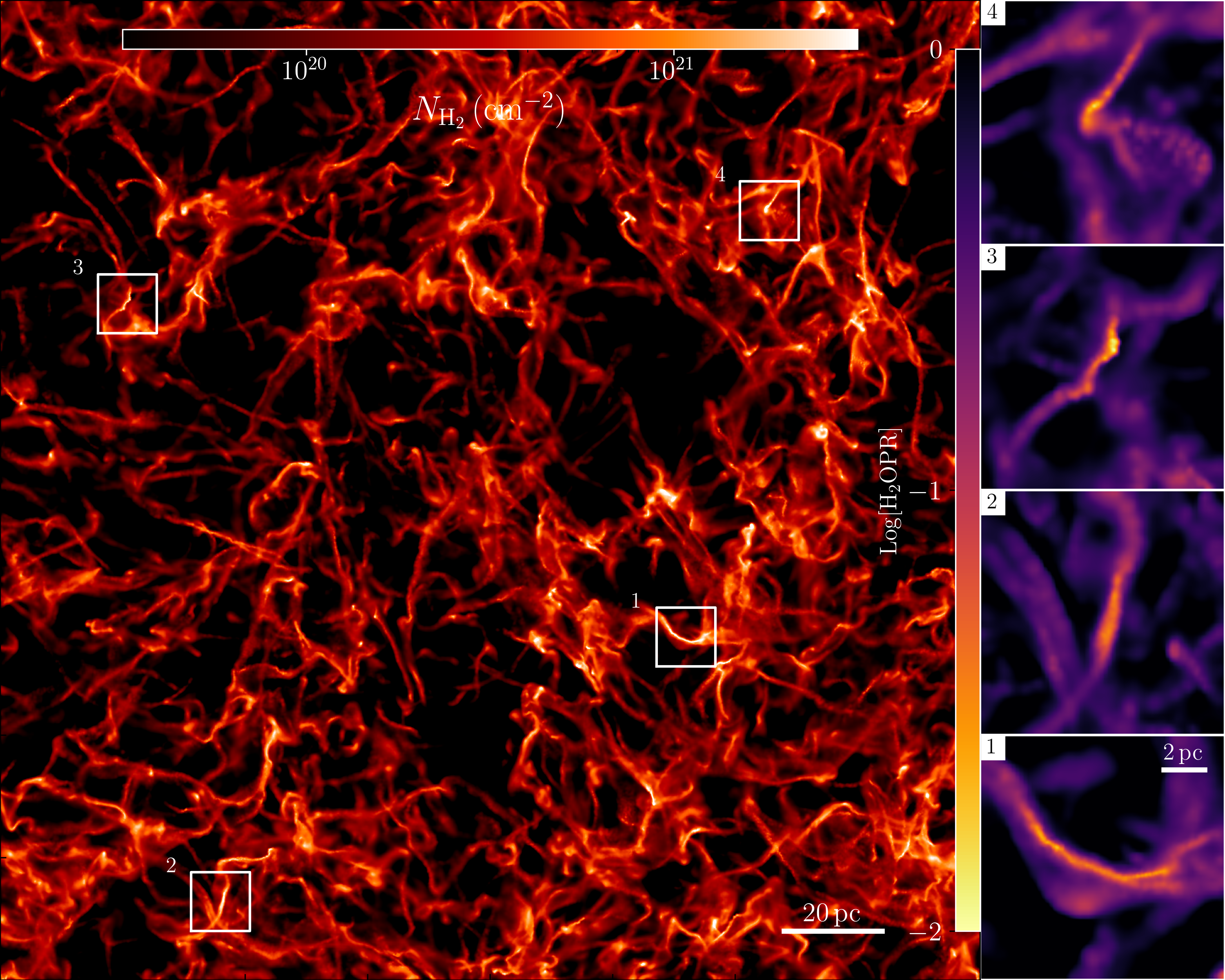}
\caption{Column density map of H$_2$ at 4~Myr. On the right, we report the OPR$^N$ in four filaments identified in the snapshot (corresponding to the four white squares in the left panel). The cloud already exhibits clear filamentary structures, and also some clumps forming within the most massive and largest ones. In these regions, the OPR is already low, with values typically below 0.1.}
\label{fig:Nmap}
\end{figure*}

More in detail, in Table~\ref{tab:fils} we report the average properties of the four selected filaments (among the most massive and extended ones, i.e.\,$M>100\,\msun$) at $t=4$ Myr. In particular, from left to right, we report the H$_2$ column density $N_{\rm H_2}$, the cosmic-ray ionisation rate $\zeta_{\rm H_2}$, the gas temperature $T_{\rm fil}$, the Mach number $\mathcal{M}$, the magnetic field $B$ magnitude, and two values for the OPR, i.e.\,OPR$^N$ and the density-weighted line-of-sight average ratio OPR$^n\equiv \langle n_{\rm o-H_2}\rangle/\langle n_{\rm p-H_2}\rangle$. Finally, in the last two columns, we report the mass per unit length $M/L$ and the filament width $W\equiv A_{\rm px}/L$, with $A_{\rm px}$ the effective pixel area of the dendrogram and $L$ the major axis length of the filament. In general, the filaments identified in our simulations show typical properties consistent with observations \citep{Arzoumanian2011}, i.e. lengths $L$ between 1 and 10 pc, axis ratios between 1:2 and 1:20, masses from a few tens up to a thousand solar masses, and, being still in an initial collapse stage, densities not exceeding $10^4\rm\, cm^{-3}$ with average temperatures around 30~K. The estimated mass per unit length ($M/L$) is compared with the critical value for collapse, obtaining a full spectrum of values ranging from $\sim 0.2$ (sub-critical) up to $\sim 3$ (super-critical), with the four reported in Table~\ref{tab:fils} lying around 1.0-1.5.

\begin{table*}
\centering
\begin{tabular}{llllllllll}
\hline

\hline

ID & $\log N_{\rm H_2}$ & $n_{\rm H_2}$ & $\log\zeta_{\rm H_2}$ & $T_{\rm fil}$ & $\mathcal{M}$ & $B$ & OPR$^{N/n}$ & $[M/L]$ & $W$\\
 & $(\rm\, cm^{-2})$ & $(\rm cm^{-3})$ &$(\rm\, s^{-1})$ & $\rm (K)$ & - & $(\mu\rm G)$ & - & $\msun pc^{-1}$ & $(\rm pc)$\\

\hline
1 & 21.28 & 768.24 & -15.53 & 31.4 & 10.60 & 7.60 & 0.10/0.06 & 77.56 & 0.10\\
2 & 21.28 & 1237.57 & -15.54 & 28.0 & 3.55 & 10.13 & 0.08/0.03 & 54.09 & 0.12\\
3 & 21.22 & 1274.69 & -15.58 & 23.9 & 7.78 & 9.92 & 0.06/0.01 & 41.33 & 0.16\\
4 & 21.38 & 2022.84 & -15.51 & 28.3 & 6.14 & 9.30 & 0.06/0.03 & 52.70 & 0.08\\
\hline

\hline
\end{tabular}
\caption{\label{tab:example} Main properties of four filaments out of hundreds in our fiducial simulation (cfr. Fig.\,\ref{fig:Nmap}) . From left to right, we report H$_2$ column density, average number density $n_{\rm H_2}$, average cosmic ray ionisation rate $\zeta_{\rm H_2}$, gas temperature $T_{\rm fil}$, Mach number $\mathcal{M}$, magnetic field magnitude $B$, OPR (using both column densities --first value-- and average number densities -- second value), mass per unit length $M/L$, and filament width $W$. All values we find are consistent with the expected ones derived from observations.}
\label{tab:fils}
\end{table*}

\begin{figure*}
\centering
\includegraphics[width=0.9\textwidth,trim=3.1cm 2.1cm 2.9cm 2.1cm,clip]{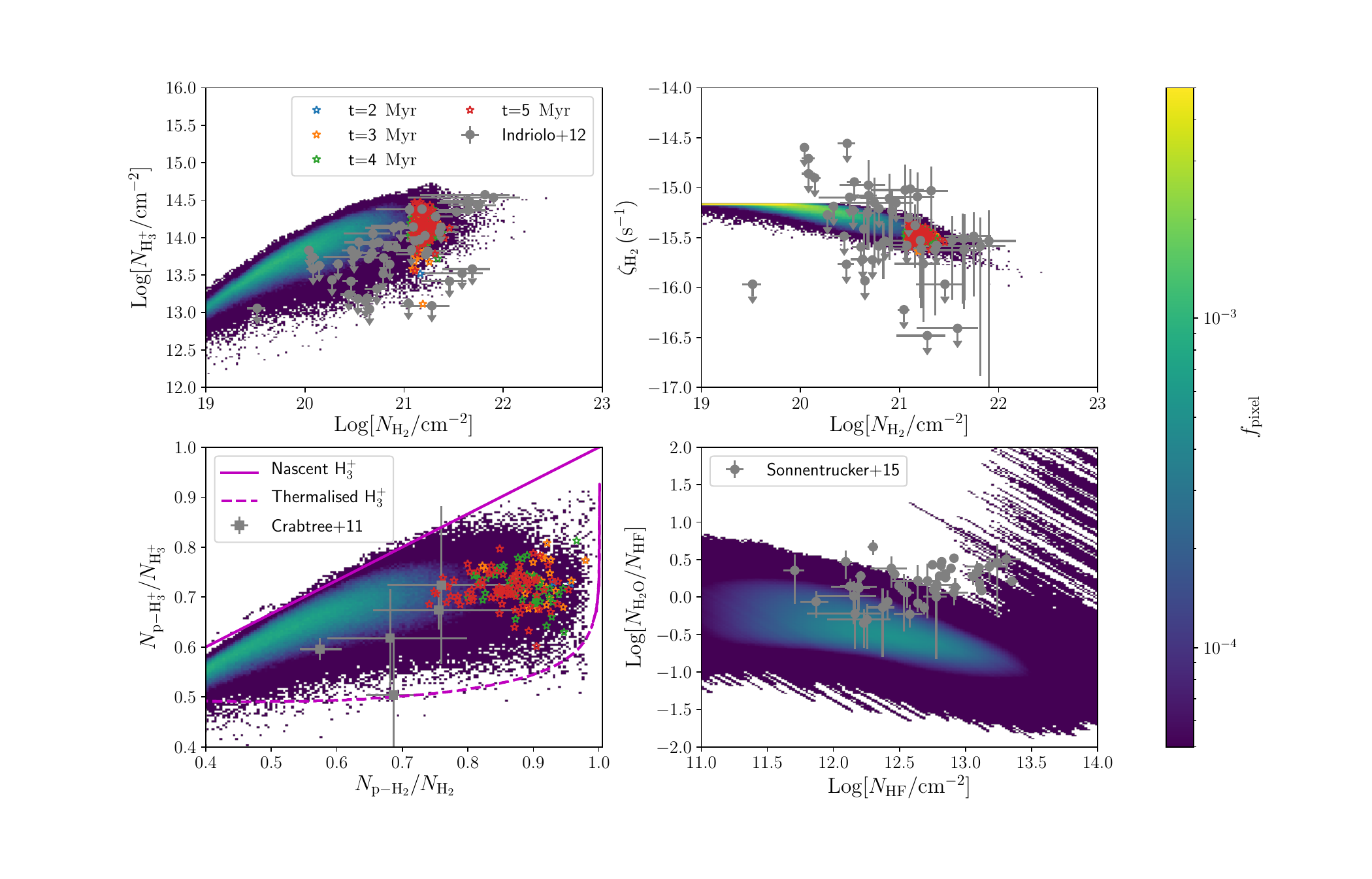}
\caption{Comparison of our simulation with existing observations of diffuse clouds, shown as grey dots \citep{Crabtree2011,Indriolo2012,Sonnentrucker2015}. The top-left panel shows the abundance of H$_3^+$ relative to H$_2$, the top-right panel the corresponding $\zeta_{\rm H_2}$, the bottom panel the para-to-total ratio for both H$_3^+$ and H$_2$, and the bottom-right panel the H$_2$O relative abundance with respect to HF, where the HF column density in our simulation is extracted assuming a rigid scaling relative to H$_2$ \citep{Indriolo2012} with scaling factors in the range $0.5\times 10^{-8}$--$3.5\times 10^{-8}$. Our simulation is in remarkable agreement with the observational results of most tracers, confirming the accuracy of our chemical and physical modelling, with only a slight discrepancy in H$_2$O, but still consistent with observations. This is likely attributed to the missing water formation channels on dust grains in our chemical network, which might be potentially important in cloud conditions \citep{Cazaux2010,Sonnentrucker2015}.}
\label{fig:fils}
\end{figure*}

To further confirm the accuracy of our modelling which, being developed with ab-initio physics, has not been calibrated to reproduce real clouds, in Fig.~\ref{fig:fils} we compare our simulated cloud 
with observations of `diffuse' clouds. In particular, we focus on properties connected to H$_2$ and its OPR, like the H$_3^+$ abundance (top-left panel) and the cosmic-ray ionisation rate $\zeta_{\rm H_2}$ estimates (top-right panel) \citep{Indriolo2012}, the para--to--total ratio of H$_3^+$ and H$_2$ \citep{Crabtree2011} (bottom-left panel), and the water--to--HF column density ratio (bottom-right panel) \citep{Sonnentrucker2015}. The coloured two-dimensional histogram represents the full distribution of pixels (0.25~pc wide) in our simulation, the grey dots are the observed data, and the blue, orange, green, and red stars the average abundances from the identified filaments at different times. Finally, the magenta lines correspond to the theoretical abundance of H$_3^+$ assuming a nascent distribution (solid line), in which the formation of H$_3^+$ from cosmic ray-induced ionisation of H$_2$ fully determines the relative abundance of the nuclear spin values, and a thermalised distribution (dashed line), in which the relative abundance is dominated by collisional exchange between H$_3^+$ and H$_2$ \citep{Crabtree2011}. The remarkable agreement suggests that our theoretical framework naturally produces reliable initial conditions for the collapsing filaments within molecular clouds (see Appendix~\ref{app:fullsuite} for a detailed analysis of the other simulations of our suite). In particular, in the bottom-left panel, both our simulation and observations lie in between the two theoretical curves, with our results more closely following the nascent distribution, which reflects the strong impact of cosmic rays. The abundance of water is slightly underestimated, particularly at high-density, likely because of the missing formation channels of water on dust grains in our network, which are potentially relevant in cold gas \citep{Cazaux2010,Sonnentrucker2015}. 

\begin{figure*}
\centering
\includegraphics[width=0.6\textwidth]{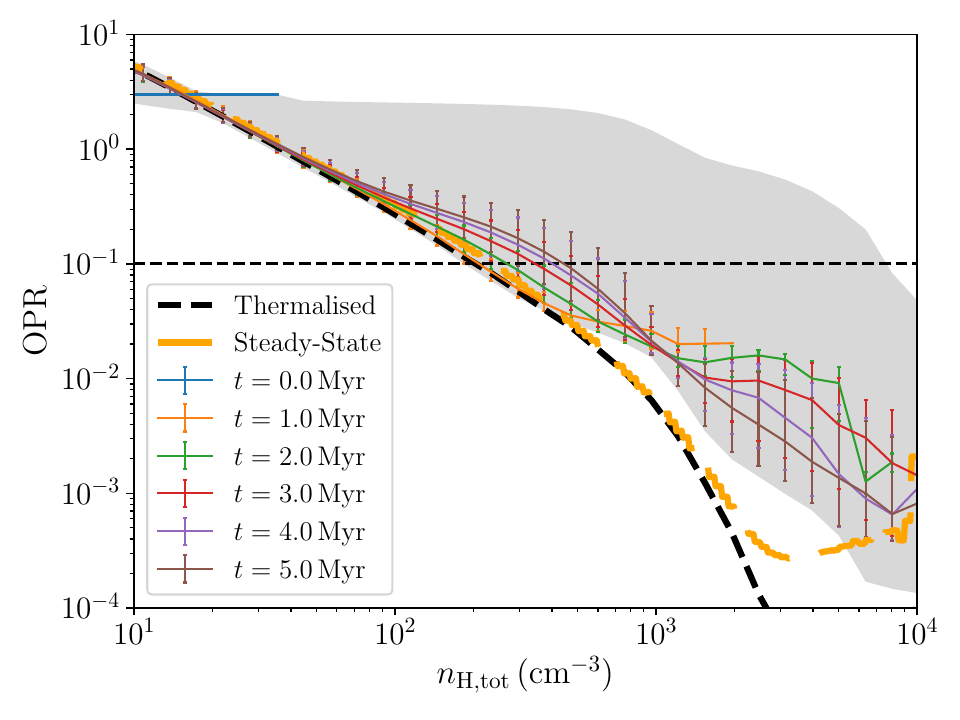}
\caption{Three-dimensional distribution of the OPR as a function of $n_{\rm H,tot}$ at different times. The grey shaded area represents the uncertainty in the physical/chemical modelling in our suite of simulations (shown for completeness), the black dashed line corresponds to the thermalised ratio, valid for temperatures below $\sim$100 K (i.e. for densities larger than 30-40 cm$^{-3}$ in our specific case), and the orange long-dashed line to the equilibrium/steady-state value. We clearly see that time plays a minor role, with the distribution being almost uniquely determined by density. Most importantly, the OPR is already well below 0.1 above $10^3\rm\, cm^{-3}$, reaching 0.001 around $10^4\rm\, cm^{-3}$. Even in the most pessimistic case, dense gas above $10^4\rm\, cm^{-3}$ never exhibits OPR values above a few 0.01. The ratio follows the thermalised/steady-state value, starting to depart around $n_{\rm H,tot}\sim 100\rm\, cm^{-3}$, when the gas becomes fully molecular. However, at high densities, the OPR settles on the equilibrium value, which is much higher than the thermalised one.}
\label{fig:opr}
\end{figure*}

In order to disentangle whether time or density play the major role in producing these results, we report
in Fig.~\ref{fig:opr} the evolution of the OPR distribution for every simulation element, as a function of the total hydrogen nuclei density. For this analysis, we directly use the local properties of the gas in the simulation, which allows us to avoid the dilution effects resulting from integrating along the line-of-sight \citep{Ferrada2021}. Each solid curve corresponds to a different time, with the error bars showing the 20th and 80th percentiles, whereas the black dashed line is the thermalised OPR, i.e. the balance between the ortho-to-para and para-to-ortho conversion, and the orange long-dashed one to the steady-state value, i.e. the ratio at chemical equilibrium self-consistently computed using our network. For completeness, we also show the total uncertainty resulting from our entire suite of simulations (see Appendix~\ref{app:fullsuite}) as a grey shaded area. 
We notice that the distribution extends to very low values ($\lesssim 0.1$) already at moderate densities ($n_{\rm H,tot}\sim 10^3-10^4\rm\, cm^{-3}$). Moreover, the distribution does not significantly change with time, suggesting that the OPR is mainly determined by density, with the dynamical evolution only being a second-order effect. This result suggests that, as soon as filaments form, the OPR is already well below 0.1,  and that higher values found for OPR$^N$ and from observations are hugely affected by dilution effects (by one order of magnitude or more).
Our results are robust even against different physical assumptions (see Appendix~\ref{app:fullsuite}), with the upper limits (grey shaded area) still exhibiting very low OPR values when $n_{\rm H,tot}\gtrsim 10^4\rm\, cm^{-3}$. At low density, the ratio tightly follows the thermalised one, which is also consistent with the equilibrium value. As soon as the gas becomes fully molecular, instead, the distribution starts to differ, but still decreases towards very small values (0.001 or less). At the highest densities probed, the OPR in the simulation settles on the equilibrium value, which showed a moderate increase above $n_{\rm H,tot}\sim 10^3\rm\, cm^{-3}$, hence departing from the thermalised one.

\section{Discussion and conclusions}
\label{sec:conclusions}
In this work, we have performed 3D MHD simulations of a molecular cloud with state-of-the-art on-the-fly non-equilibrium chemistry, finding that filaments are characterised by already low OPR values. Our simulations show a remarkable agreement with available observations, without any \emph{a priori} tuning of the model. It is important to remark that a proper comparison with other available works is difficult, as direct measurements or estimates of the OPR in diffuse and molecular clouds are rare. In particular, the few data in diffuse clouds \citep{Crabtree2011} suggest values around 0.3-0.7, slightly higher than ours, but in relatively good agreement when considering the observational uncertainties (e.g.\,optical thickness of these lines and the possible dilution effects along the line-of-sight). Some indirect estimates in between 0.001-0.2 have also been provided for pre-stellar cores \citep{Trompscot2009,Brunken2014,Pagani2013}, i.e.\,in regions with densities higher than those explored by our simulations and representing an advanced stage of the star-formation process. However, the strong assumptions made to infer this fundamental quantity from different chemical proxies do not allow to get a reliable estimation from observations.
In this context, our study represents a relevant step forward to improve the fundamental knowledge of the star-formation process.

The absence of sulfur chemistry in our work might represent a limitation, since it could reduce the ortho-to-para conversion efficiency by removing H$^+$ via the reaction path \mbox{S + H$^+ \to$ S$^+$ + H} \citep{Furuya2015}. However, \citet{Furuya2015} show that (i) sulfur is quickly adsorbed on dust grains as soon as S$^+$ recombines and, as a consequence, (ii) this effect is relevant only for extremely high metal abundances, which are not compatible with the measured values of sulfur from observations. 
Hence, we assume that sulfur chemistry does not play a relevant role within the context of our model, and therefore it does not affect our conclusions.

Concluding, our state-of-the-art three-dimensional magneto-hydrodynamic simulations of molecular clouds formation with on-the-fly non-equilibrium chemistry indicate that the H$_2$ ortho-to-para ratio quickly evolves with density, reaching very low values (but far from the thermalised one) already at moderate densities typical of proto-filaments. This is particularly relevant for smaller scale studies, in which the unconstrained initial OPR is either varied in the allowed range to bracket its effect on the results \citep[see, e.g.][]{Sipila2015,Kong2015,Bovino2020a} or conservatively assumed to be around 0.1 \citep[see, e.g.][]{Jensen2021}, i.e.\,larger than ours ($10^{-3}-10^{-2}$). The results in this work therefore establish that the initial conditions of the star formation process in filaments are characterised by already low OPR and typically high $\zeta_\mathrm{H_2}$, both conditions that would dramatically boost the deuteration mechanism and shorten the corresponding timescales. We notice that, even in filaments  characterised by low cosmic ray ionisation rates \citep{Indriolo2012}, the expected OPR would be only moderately higher, and never as high as 0.1.
For the first time, we have been able to determine the initial conditions of star-forming filaments from ab-initio conditions, in particular the chemical abundances of important species like H$_2$ (ortho- and para-), H$_3^+$, CO, and H$_2$O, which have far-reaching implications for the deuteration process (hence on the reliability of chemical clocks), for the observed HDO/H$_2$O ratios in planet-forming regions (which strongly depends on the OPR), and its connection with the origin of water in our Solar System.

 \begin{acknowledgements}
 AL acknowledges funding from MIUR under the grant PRIN 2017-MB8AEZ.
 The computations/simulations were performed with resources provided by the Kultrun Astronomy Hybrid Cluster. This research made use of \textsc{astrodendro}, a \textsc{Python} package to compute dendrograms of astronomical data (\url{http://www.dendrograms.org}), and \textsc{pynbody}, a \textsc{Python} package to analyse astrophysical simulations \citep{pynbody}. Part of this work was supported by the German \emph{Deut\-sche For\-schungs\-ge\-mein\-schaft, DFG\/} project number Ts17/2--1.
\end{acknowledgements}

\bibliographystyle{aa}
\bibliography{mybib_D}

\begin{appendix}
\section{Numerical methods}\label{app:methods}
\subsection{Turbulence driving}
In order to drive turbulence in the box with properties similar to observed molecular clouds, the gas distribution is stirred by a random acceleration field obtained via an Ornstein–Uhlenbeck process \citep{Federrath2010,Bauer2012}. The energy power spectrum normalisation is set to obtain a velocity dispersion in the cloud $\sigma_{v} \sim 7\,\kms$, and the auto-correlation time of the random process is set to $\tau = 10$ Myr. Energy is injected only at large scales, according to a parabolic power spectrum peaking at $k=3$ (extending from $k=2$ up to $k=4$), assuming a half-solenoidal half-compressive driving \citep{Bauer2012}. This assumption is reasonably consistent with recent simulations of molecular cloud formation in which turbulence was driven self-consistently via random supernova explosions \citep{Padoan2016}. During this relaxation phase, which lasts for 50~Myr in order to let turbulence fully develop, we do not include self-gravity, and we assume an isothermal equation of state with $T=5000$~K and no chemistry evolution \citep{Mandal2020}. This allows us to start the self-consistent evolution of the cloud with ab-initio chemical abundances typical of the warm neutral medium, avoiding any pre-processing of the species during the relaxation. Nevertheless, to further corroborate our results and isolate the effect of self-gravity, we also perform an additional experiment in which chemistry and proper cooling are already included during the relaxation phase.

\subsection{Sink formation}
Near the end of the simulation, we expect clumps to form in filaments and a few resolution elements to reach very high densities, thus requiring very short time-scales to integrate the dynamics.  In order to avoid this undesired slow-down and let the simulation to evolve for longer times, we convert the gas hitting the resolution limit of the simulation into proto-stellar objects, commonly dubbed `sink particles', which are able to grow via accretion of surrounding gas. In this work, we include a simple sink formation scheme, i.e. gas particles i) above $n_{\rm H,tot}>10^{10}\rm\, cm^{-3}$ ii) showing negative velocity divergence and iii) located at a relative gravitational potential minimum are converted into sink particles, unless another sink is found within their kernel volume. After a sink has formed, we allow it to accrete gas within its kernel volume, corresponding to a sphere enclosing an effective number of 32 neighbours that matches the sink formation conditions. However, since our main interest is in the pre-stellar phase, rather than in the star formation process itself, we stop our simulations after a few sinks have formed in the box, making the details of the sink formation scheme almost irrelevant.

\subsection{Cosmic ray flux determination}
 The cosmic ray ionisation rate in molecular clouds is affected by several uncertainties, since it strongly depends on the physical conditions inside and outside the cloud. For this reason, most studies to date explore different values of the cosmic ray ionisation rate $\zeta_{\rm H_2}$ typically in the range $1.3\times 10^{-17}-1.3\times 10^{-16}$. While\st{,} in some of our simulations we also employ a constant value, in a real cloud cosmic rays are attenuated as they move deeper into it, hence a more consistent modelling should take into account this aspect. However, detailed cosmic ray propagation in 3D simulations is computationally expensive, and would add an additional layer of complexity to our already computationally expensive simulations. For this reason, in this work we opt for an effective model which, starting from the local properties of each resolution element, allows us to follow the variations of $\zeta_{\rm H_2}$ in the cloud, at a moderate computational cost\citep{Padovani2018}:
\begin{equation}
    \zeta_{\rm H_2}=\zeta_{\rm H_2,p}+\zeta_{\rm H_2,e},
\end{equation}
where the two contributions are from protons and electrons, respectively, and are defined as
\begin{eqnarray}
    \zeta_{\rm H_2,p} &=&
    \left\{\begin{array}{cc}
         6.8\times10^{-16}N_{20}^{-0.423} & N_{\rm eff}<10^{25}\rm cm^{-2} \\
         5.4\times10^{-18}\exp(-\Sigma_{\rm eff}/38) & {\rm otherwise}
    \end{array}\right.
    \\
    \zeta_{\rm H_2,e} &=&
    \left\{\begin{array}{cc}
         1.4\times 10^{-19}N_{20}^{-0.04} & N_{\rm eff}<10^{25}\rm cm^{-2} \\
         3.3\times 10^{-20}\exp(-\Sigma_{\rm eff}/71)& {\rm otherwise}
    \end{array}\right.
\end{eqnarray}
where $N_{20}=N_{\rm eff}/10^{20}\rm\, cm^{-2}$ and $\Sigma_{\rm eff}=2.36m_{\rm H}N_{\rm eff}$, with $N_{\rm eff}$ the effective column density traversed by cosmic rays. To determine $N_{\rm eff}$, we assume that the magnetic field lines are relatively not curved (as expected in this earlier star-formation stage) and they have small intensity variations, allowing us to consider the total column density $N_{\rm H_2}=N_{\rm o-H_2}+N_{\rm p-H_2}$ as a reliable first-order approximation, which we determine using a local approximation \citep{grassi17} in which $N_x = 1.87\times 10^{21}(n_x/10^3\rm\, cm^{-3})^{2/3}$, of $N_{\rm eff}$. When the magnetic field lines are not straight, our estimate of $N_{\rm eff}$ represents a lower limit to the actual value, which translates into our cosmic ray ionisation rate being an upper limit. This is why, in our suite, we chose an average model among those in the literature \citep{Padovani2009}, and also explored very different conditions, i.e. a conservative and uniform $\zeta_{\rm H_2}=1.3\times 10^{-17}$ s$^{-1}$, and the locally varying one, to bracket the possible conditions of real clouds.

\subsection{Ortho-to-para H$_2$ conversion on dust}
Recent experimental works both on amorphous solid water\citep{Ueta2016} as well as on bare silicates\citep{Tsuge2021}, show the efficiency of the ortho-to-para H$_2$ conversion on the surface of dust grains. To consistently follow this process within our simulations and evaluate its overall impact on the OPR, we employ state-of-the-art frameworks\citep{Bovino2017,Furuya2019}, in which the conversion rates in units of s$^{-1}$ are defined as
\begin{eqnarray}
    k_\mathrm{op} = k_\mathrm{ads}^\mathrm{oH_2} \eta_\mathrm{op}\\
    k_\mathrm{po} = k_\mathrm{ads}^\mathrm{pH_2} \eta_\mathrm{po}
\end{eqnarray}
where the $k_\mathrm{ads}^i = Sv_i\sigma_\mathrm{dust}$ are the adsorption rates of the species on the surface of grains, with $S$ being the sticking coefficient (here assumed to be 1 for simplicity), $v_i$ the thermal gas speed, and $\sigma_\mathrm{dust}$ the distribution-averaged grain geometrical cross-section. The efficiency of the process is regulated by the factor $\eta_i$ which represents the competition between the conversion process and desorption, and it is defined as\citep{Furuya2019}
\begin{eqnarray}
    \eta_\mathrm{op} = \frac{t_\mathrm{des}}{t_\mathrm{des}+\tau_\mathrm{conv}}\frac{1}{1+\gamma}\\
    \eta_\mathrm{po} = \frac{t_\mathrm{des}}{t_\mathrm{des}+\tau_\mathrm{conv}}\frac{\gamma}{1+\gamma}
\end{eqnarray}
where the desorption time is calculated as the minimum between the thermal desorption time and the cosmic-ray induced desorption time, $\tau_{\rm conv}$ is the experimental conversion time\citep{Tsuge2021}, fitted as $\tau_\mathrm{conv} = 6.3\times 10^4 T_\mathrm{dust}^{-1.9}$, and $\gamma = 9 \exp{(-170.5/T_\mathrm{dust})}$ is the thermalised value of the ortho-to-para H$_2$ ratio, assuming the energy difference between o‐H$_2$ and p‐H$_2$ on the grains is the
same as that in the gas phase.
This allows us to include in the H$_2$ rate equations both the ortho-to-para conversion and its inverse process (para-to-ortho). However, since the inverse process is in general negligible at low temperatures\citep{Bovino2017}, we opt for completely neglecting it in our simulations, which in practice corresponds to assuming $\gamma=0$ \citep{Bovino2017}. We note that our approach is based on a single average binding energy approximation, in particular $E_b = 600$ K as reported in literature\citep{Perets2006,Vidali2010,He2014}, and refer to other recent works\citep{Furuya2019} for a more comprehensive treatment which also includes the thermal hopping between different adsorption sites. The effect of these processes on the evolution of the H$_2$ OPR is discussed in Appendix~\ref{app:fullsuite}.

\subsection{HF abundance in diffuse clouds}
Observations of water in diffuse clouds typically lack a direct measure of the H$_2$ column density, using instead HF (or other hydrides) as an alternative proxy\citep{Sonnentrucker2015}. Despite the correlation $N_{\rm HF}=\chi_{\rm HF}N_{\rm H_2}$ is found to be quite tight, the uncertainty in the conversion factor reaches up to a factor of $\sim 7$, ranging from $0.5\times 10^{-8}$ up to $3.5\times 10^{-8}$ \citep{Indriolo2012}. For this reason, and considering the fact that our chemical network does not include HF, no simple comparison between simulations and observations exists, and particular attention must be taken when converting H$_2$ to HF (or viceversa). For the comparison between our runs and observations of Fig.~\ref{fig:fils}, we opt therefore for a more sophisticate procedure, i.e. for each value of $N_{\rm H_2}$ in our simulations, we estimate the $N_{\rm HF}$ using 100 different values of $\chi_{\rm HF}$ within the observed range, so that this source of uncertainty is properly accounted for, and combine all these measures in a single 2D histogram reported in the bottom-right panel of the figure, appropriately normalised to the total number of measures available.

\section{Relaxation phase}
\label{app:relax}
\begin{figure}
\centering
\includegraphics[width=\columnwidth,trim=0cm 1cm 1cm 1cm,clip]{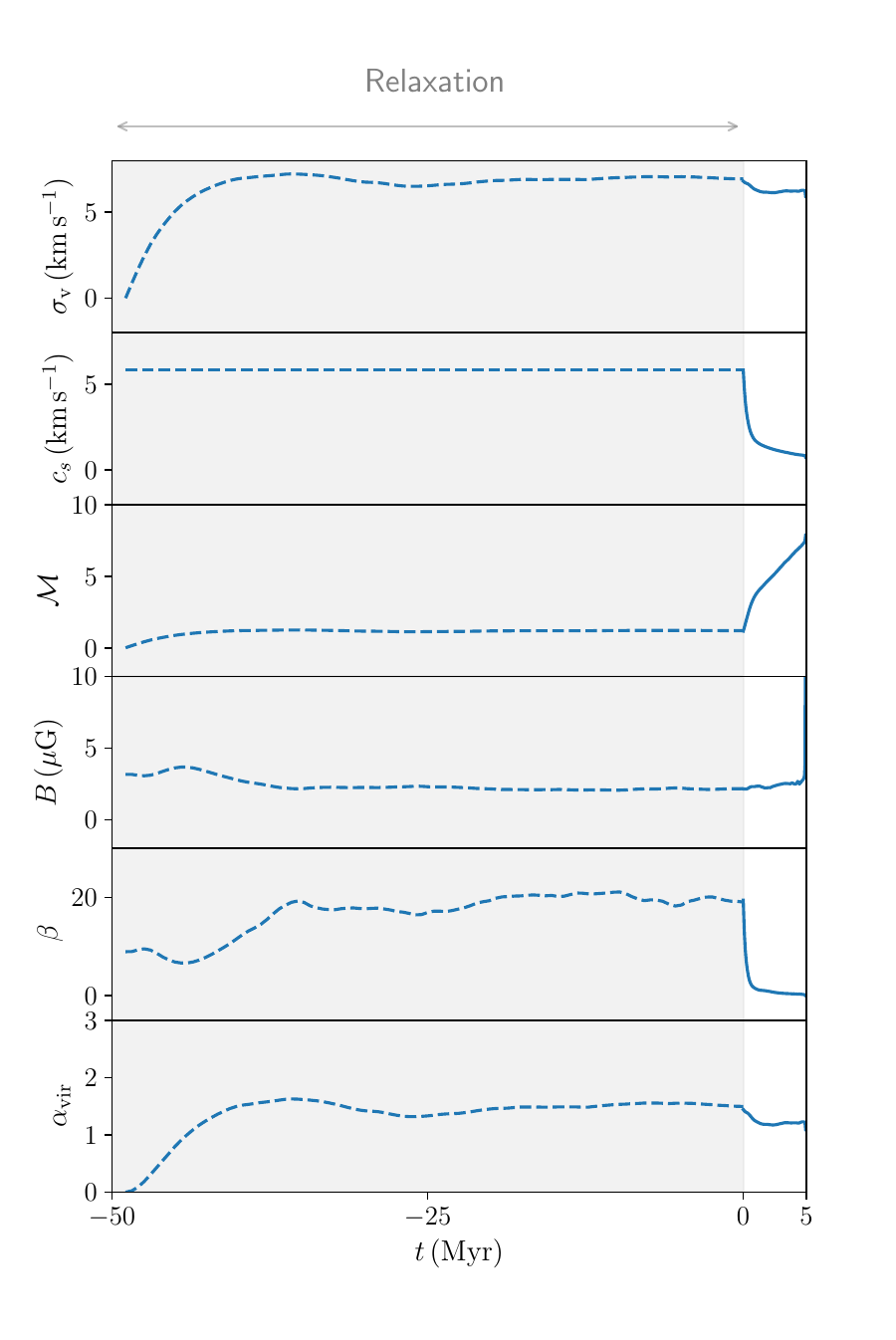}
\caption{Evolution of the global properties of the turbulent cloud in our fiducial model, during the relaxation phase (shaded area with negative times) and the actual run (positive times). While turbulence leads to an initial increase of $\sigma_{\rm v}$ (reflected in all the other properties it affects), the sound speed remains constant during relaxation, producing a Mach number slightly above unity. During the run, instead, $\mathcal{M}$ increases quickly to the typically observed values, because of cooling. On average, the magnetic field does not change significantly, as long as the typical density interval remains small.}
\label{fig:props}
\end{figure}

 Our simulation suite is composed of four runs, three of them starting from our fiducial isothermal relaxation without chemistry, that we call \textit{RelaxIso} and one in which proper cooling and chemistry are accounted for also during relaxation, that we call \textit{RelaxChem}, in which $\zeta_{\rm H_2}=1.3\times 10^{-17}$ s$^{-1}$ is kept constant over time in the entire box. The relaxation phase is necessary to guarantee that turbulence fully develops in the box, producing initial conditions of the actual simulations that more closely represent realistic molecular clouds\citep{Federrath10}. 
 To give an idea of the global evolution of our fiducial simulated cloud, we show in Fig.~\ref{fig:props} the main properties of the box during both the relaxation phase (reported as a grey shaded area with negative times) and the actual run (reported with positive times), i.e. the density-weighted average 3D velocity dispersion $\sigma_{\rm v}$, sound speed, Mach number, $\beta=P/P_{\rm B}$ parameter, with $P=\rho c_s^2$ the thermal pressure (with $\rho$ the total gas density) and $P_{\rm B}=\langle B\rangle^2/(8\pi)$ the magnetic pressure, and virial parameter $\alpha_{\rm vir}=5\sigma_{\rm v}^2L/(3GM)$, with $L$ the box size and $M$ the box total mass. During the relaxation phase, $\sigma_{\rm v}$ increases up to the desired value as a result of turbulence driving, whereas the sound speed stays constant, because of the constant temperature assumption, which also reflect in the virial parameter increase to about 1.5 and the Mach number staying close to unity. The modest change in density distribution results in a small decrease of the average magnetic field, and the corresponding increase of $\beta$ by a factor of two.
 After relaxation, $\sigma_{\rm v}$ only modestly varies, while $c_s$ significantly decreases because of cooling, producing a rapid increase of $\mathcal{M}$ to the typically observed values. These variations reflect directly on $\beta$ and $\alpha_{\rm vir}$, except for the magnetic field that suddenly rises above 10~$\mu$G around $t=5$~Myr, when a large portion of the gas starts to collapse.

  \begin{table*}
\centering
\begin{tabular}{llllllll}
\hline

\hline

Model & $\langle n_{\rm H,tot}\rangle$ & $\langle T\rangle$ & $\langle \sigma_{v}\rangle$ & $c_s$ & $\mathcal{M}$ & $\langle \beta\rangle $ & $\alpha_{\rm vir}$\\
 & $(\rm cm^{-3})$ & $(\rm K)$ & $(\rm km s^{-1})$ & $(\rm km s^{-1})$ & - & - & -\\ 

\hline
RelaxChem & $123.8$ & $128$ & $7.525$ & $0.663$ & $11.35$ & $0.318$ & $1.761$\\
RelaxIso & $6.2$ & $5\times 10^3$ & $7.162$ & $5.820$ & $1.230$ & $2.008$ & $1.600$\\
\hline

\hline
\end{tabular}
\caption{Main properties of our turbulent box after 50~Myr of relaxation for the two different considered models \textit{RelaxIso} and \textit{RelaxChem}. The inclusion of cooling and chemistry during the relaxation phase allows the gas to spread over a larger density/temperature interval, resulting in very different thermal properties (typical of colder and denser gas), whereas those fully determined by turbulence remain almost identical (the 3D velocity dispersion $\sigma_{\rm v}$ and the virial parameter $\alpha_{\rm vir}$).}
\label{tab:relax}
\end{table*}
 
 \begin{figure*}
\centering
\includegraphics[width=0.75\textwidth,trim=1cm 1.5cm 2cm 1.5cm, clip]{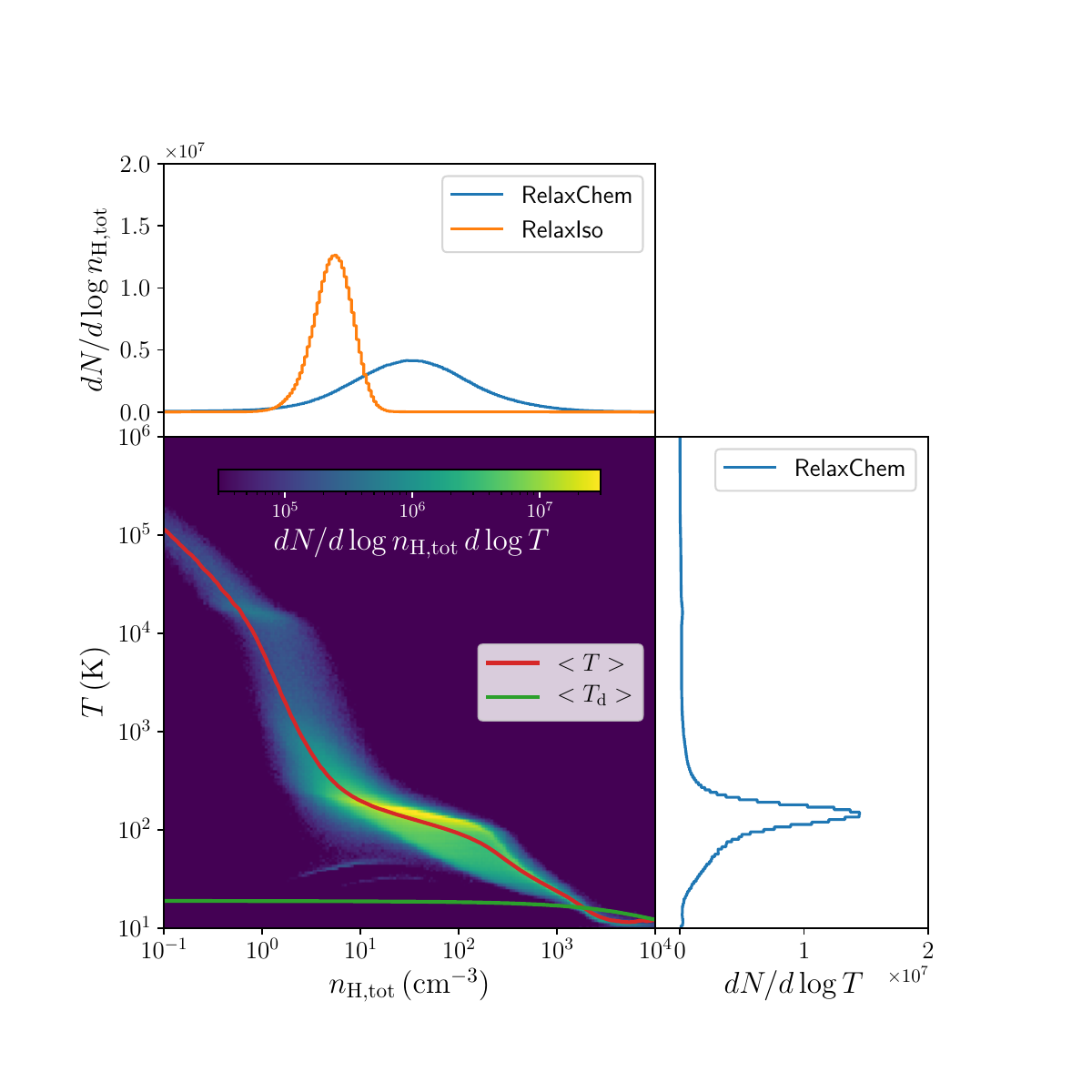}
\caption{Distribution of the gas at the end of the relaxation phase. The density--temperature diagram corresponds to \textit{RelaxChem} (notice that this distribution does not change when gravity is included, apart from extending to higher densities as the gas collapses), with the red line representing the average temperature, and the green one the average dust temperature. The temperature-only distribution in the right-hand panel shows the typical temperature of the gas, 100-200~K, higher than the typical temperature assumed in isothermal molecular cloud simulations. In the top panel, we show instead the density distribution for both relaxation models, where \textit{RelaxChem}, extending to higher densities because of cooling, is shown in blue, and \textit{RelaxIso}, more concentrated around the initial density, in orange. }
\label{fig:pdf}
\end{figure*}
 
 For completeness, we also report in Table~\ref{tab:relax} the same main properties at the end of the relaxation phase for the two relaxation models we considered, in addition to the density-weighted average density and temperature. While the turbulence-driven properties are the same in both relaxation models, the thermodynamic is not, resulting in very different average temperatures and densities. To better clarify how the gas is distributed in the two cases, in Fig.~\ref{fig:pdf} we show the density and temperature distributions (the latter only for the \textit{RelaxChem} case). Overlaid on the density-temperature plot of \textit{RelaxChem} we also report the average temperature (red) and dust temperature (green) curves, that differ significantly at low density, while couple at $n_{\rm H,tot}\sim 10^4\rm\, cm^{-3}$. In the right panel, we show the temperature-only distribution, which peaks around 100--200~K, at which most of the `diffuse' gas settles. In the top panel we report instead the density-only distribution, in this case also for \textit{RelaxIso} (orange histogram). As expected, in \textit{RelaxChem}, gas cooling allows the gas to get denser after turbulence-induced compression, spreading over a larger density interval, that follows a Gaussian-like profile (the typically expected Log-Normal density probability distribution function). On the other hand, \textit{RelaxIso} keeps the gas to moderate densities, producing a much narrower distribution centred around the average density of the initial conditions.

 \section{The full simulation suite}
 \label{app:fullsuite}
 
\begin{figure*}
\centering
\includegraphics[width=\textwidth,trim=3.1cm 2.1cm 2.9cm 2.1cm,clip]{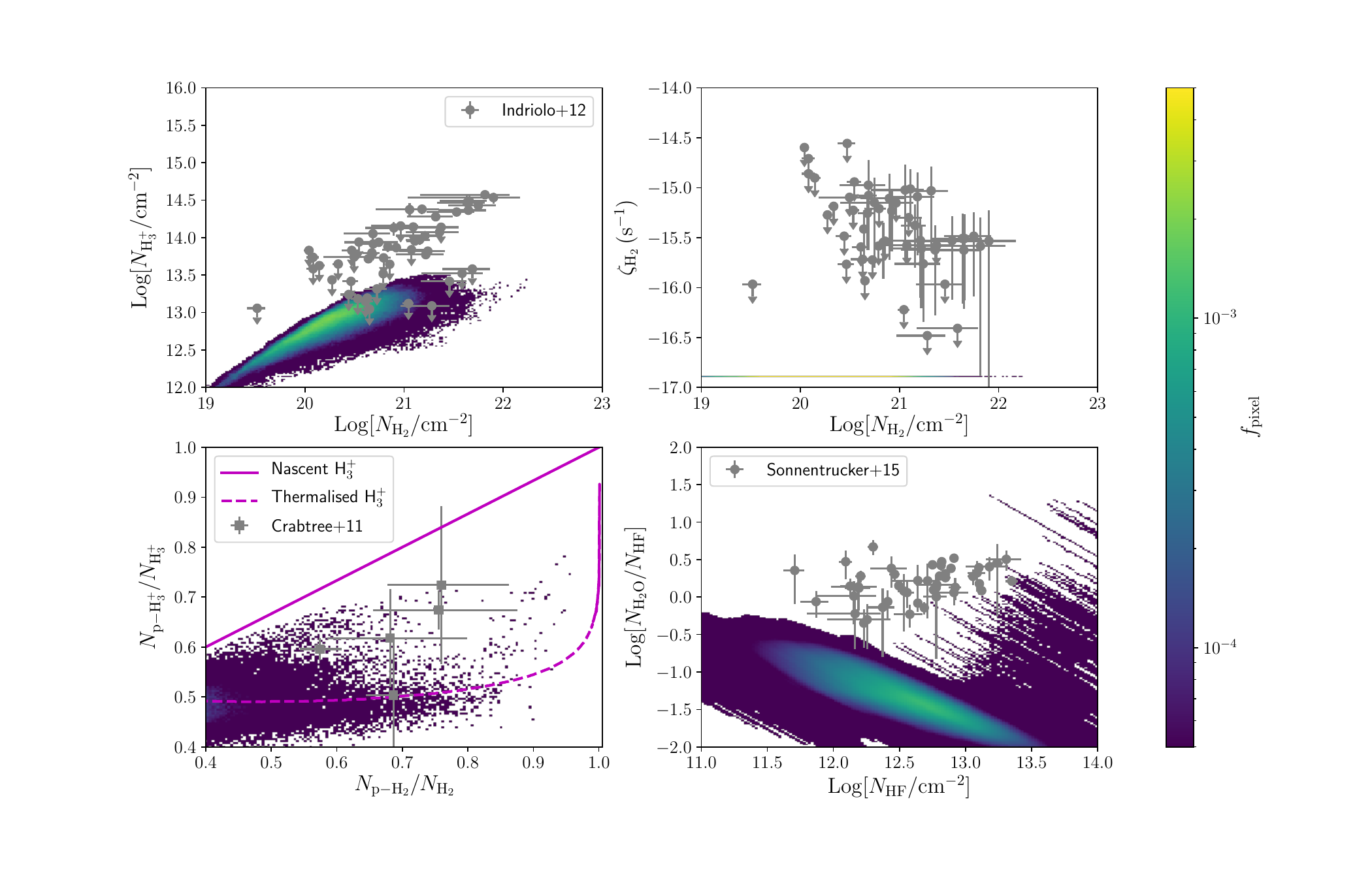}
\caption{Same as Fig.~\ref{fig:fils} for model \textit{RelaxIso\_SG}. Unlike in our fiducial model, here the abundances are much lower than those observed, consistently with the huge difference in $\zeta_{\rm H_2}$ between the observationally-inferred value and the one in the simulation. In this case, the data in the bottom panel are in better agreement with the thermalised distribution, because of the small effect of cosmic ray-induced ionisation of H$_2$ relative to atom exchange reactions.}
\label{fig:obs_crfix}
\end{figure*}

\begin{figure}
\centering
\includegraphics[width=0.95\columnwidth,trim=0cm 0cm 0cm 0.3cm, clip]{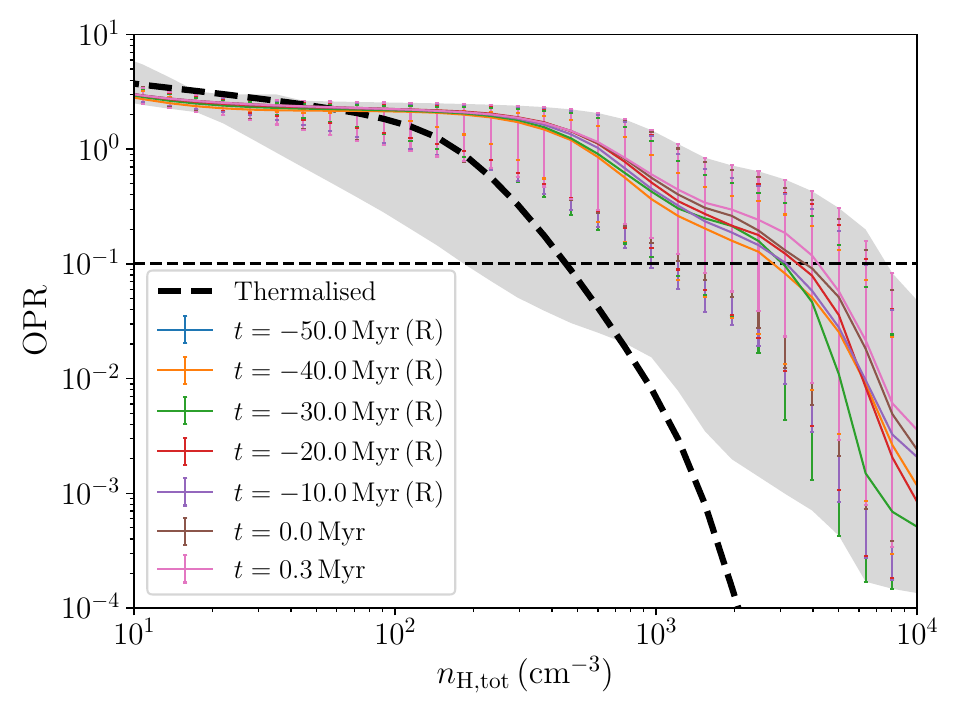}
\includegraphics[width=0.95\columnwidth,trim=0cm 0cm 0cm 0.3cm, clip]{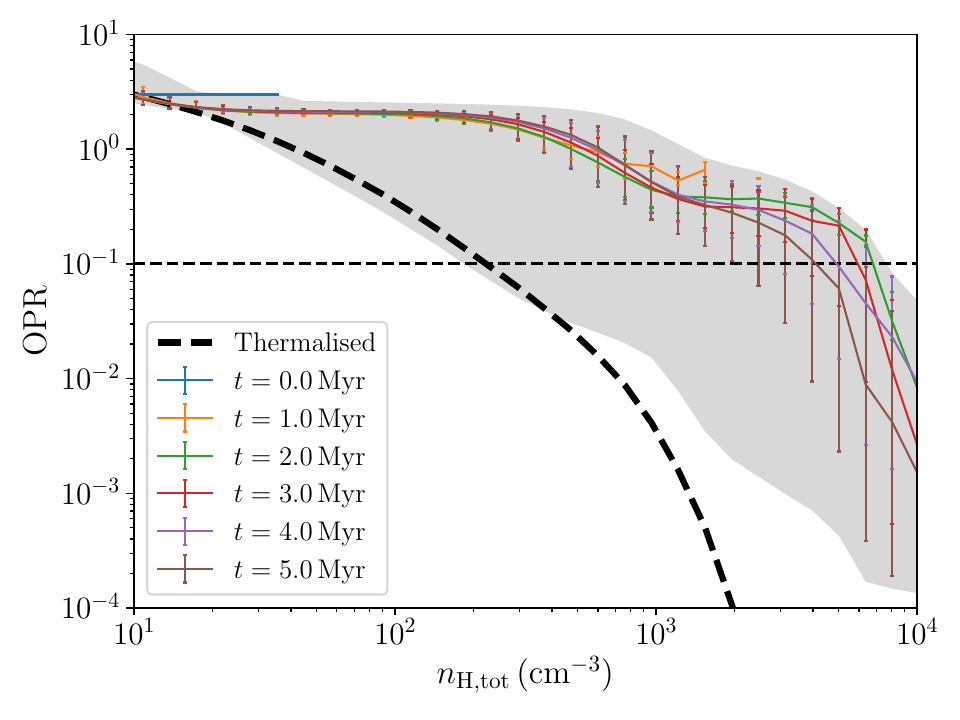}\\
\includegraphics[width=0.95\columnwidth,trim=0cm 0cm 0cm 0.3cm, clip]{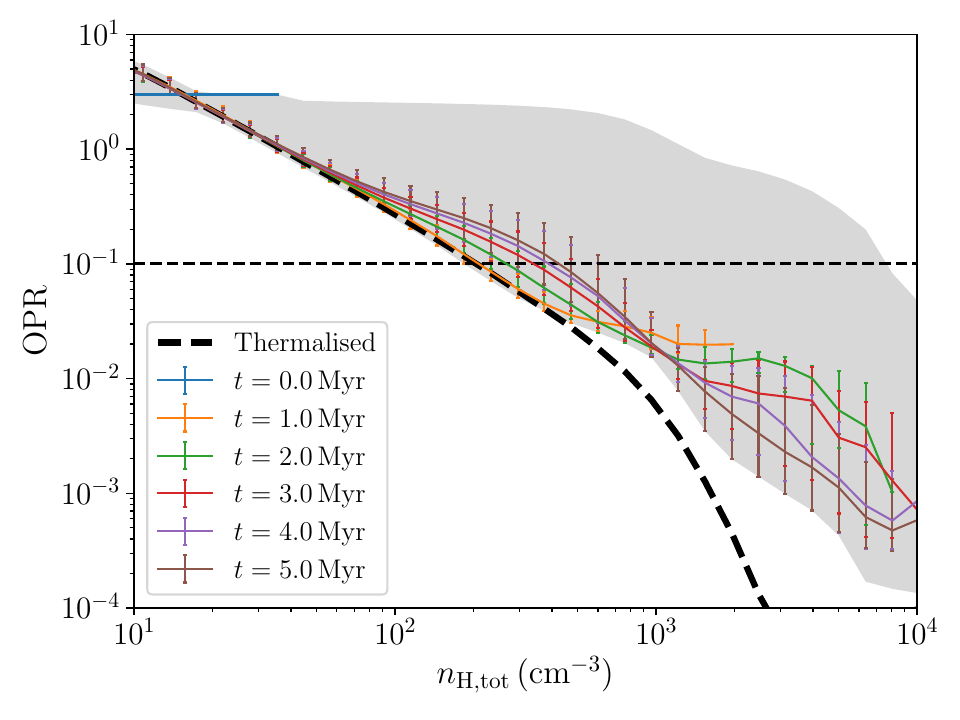}\\
\caption{Same as Fig.~\ref{fig:opr} for models \textit{RelaxChem\_SG} (top panel), \textit{RelaxIso\_SG} (middle panel), and \textit{RelaxIso\_SG\_OPdust} (bottom panel). The OPR distribution shows very mild variations with time in all cases but, while the bottom panel shows almost no difference with the fiducial model, the other two runs are typically offset upwards, a result that reflects the very low cosmic ray ionisation rate, not typical of molecular cloud conditions. Nevertheless, the OPR at $n_{\rm H,tot}\sim 10^4\rm\, cm^{-3}$ is always well below 0.1.}
\label{fig:opr_ext}
\end{figure}

\begin{figure}
\centering
\includegraphics[width=0.95\columnwidth,trim=0.5cm 0cm 0cm 0.3cm, clip]{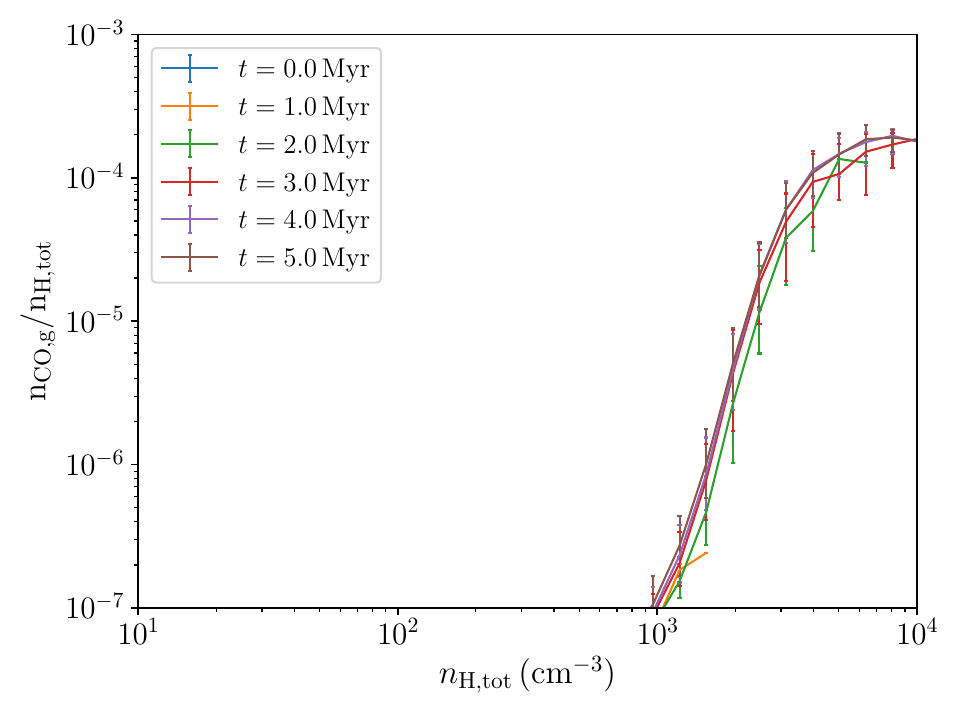}
\caption{Fraction of CO in gas phase in our fiducial model, as a function of total hydrogen density. Despite the high $\zeta_{\rm H_2}$, CO is able to form efficiently, reaching the canonical abundance in gas above $n_{\rm H,tot}=3-4\times 10^3\rm\, cm^{-3}$. We also notice freeze-out on dust grains starting to deplete CO at $n_{\rm H,tot}\sim 10^4\rm\, cm^{-3}$, as expected, although the number of resolution elements above this density is very low. We also find that the abundance does not significantly vary with time, similarly to the OPR, as long as depletion remains almost negligible.}
\label{fig:COab}
\end{figure}

 The four simulations we performed are meant to cover most of the plausible parameter space, thus properly constraining our models relative to observations. In particular, compared to our fiducial model, we consider: a) \textit{RelaxChem} plus self-gravity, named \textit{RelaxChem\_SG}, b) \textit{RelaxIso} plus self-gravity, named \textit{RelaxIso\_SG}, c) the fiducial model (see Main Text), and d) the fiducial model with the addition of ortho--to--para conversion on dust (as described in Section~\ref{app:methods}), named \textit{RelaxIso\_CR\_OPdust}. Similar to Fig.~\ref{fig:opr} in the Main Text, Fig.~\ref{fig:opr_ext} reports the OPR evolution for models \textit{RelaxChem\_SG} (top panel), \textit{RelaxIso\_SG} (middle panel), and \textit{RelaxIso\_SG\_OPdust} (bottom panel) for completeness. \textit{RelaxIso\_CR\_OPdust} gives almost identical results to our fiducial model, suggesting that the OPR conversion on dust, which slightly accelerates the ortho-to-para conversion, does not change our picture significantly, and that the gas-phase collisions with H$^+$ and H$_3^+$ are already efficient enough to bring the OPR down, without the need of this additional mechanism. 
 For \textit{RelaxChem\_SG}, in which chemistry is also evolved during the relaxation phase, the OPR distribution in these stages is reported with negative times and an additional `(R)' in the label. We immediately see that, even without self-gravity, when cooling is included, the denser gas stirred by turbulence quickly settles on a steady-state OPR distribution, and the addition of gravity\footnote{Notice that, in this last case, dense filaments and clumps already form during the relaxation phase, and they collapse in less than 1~Myr after self-gravity is included.} does not alter the distribution at all. Compared to the fiducial run, here the OPR is typically higher, farther from the thermalised value. Nevertheless, a clear drop can be observed around $n_{\rm H,tot}\sim 3-4\times 10^3\rm\, cm^{-3}$, which also in this case yields an OPR well below 0.1 at the typical densities of star-forming filaments ($n_{\rm H,tot}\gtrsim 10^4\rm\, cm^{-3}$). A similar trend can be observed in \textit{RelaxIso\_SG}, although the high density drop is even steeper that in the previous case, consistent with the weak time-dependence of the distribution (notice indeed that in this second case chemistry was not present during the initial relaxation, hence the processing time is much shorter). Combining the results of these two models, we can conclude that the higher OPR distribution relative to our fiducial model is the result of the very low cosmic ray ionisation rate, which is reasonable for protostellar cores, but not for the typical conditions of molecular clouds\citep{Padovani2009,Padovani2018}. Nonetheless, all our models, even the most pessimistic ones, result in low OPR ($\lesssim 0.01)$ at the typical densities of star-forming filaments, and this further corroborates our conclusions in the Main Text.
 To further support our claims about the effect of a too low $\zeta_{\rm H_2}$, we compare in Fig.~\ref{fig:obs_crfix} our \textit{RelaxIso\_SG} with observations, as in Fig.~\ref{fig:fils}. We immediately see that, with respect to our fiducial run, the results here are completely off, with the simulation yielding lower abundances than those observed. This is perfectly consistent with the discrepancy between our assumed $\zeta_{\rm H_2}$ with respect to the observationally-inferred value \citep{Indriolo2012}, even though these results might be compatible with the claimed non-detections. Moreover, in the bottom-left panel, our results almost perfectly lie on the thermalised distribution, consistently with the fact that cosmic ray-induced ionisation of H$_2$ becomes almost negligible with respect to the atom exchange reactions dominating in thermalised conditions.
 
 Despite not being the focus of this study, we report in Fig.~\ref{fig:COab} the abundance of gaseous CO in our fiducial simulation for completeness at different times. CO forms efficiently in our filaments, reaching the canonical fraction of about $10^{-4}$, despite the high cosmic ray ionisation rate. We also notice that CO does not exhibit any strong time-dependence in the probed density range, but this is not surprising since freeze-out on dust grains still has a negligible impact.

 \begin{figure*}[ht]
\centering
\includegraphics[width=\textwidth]{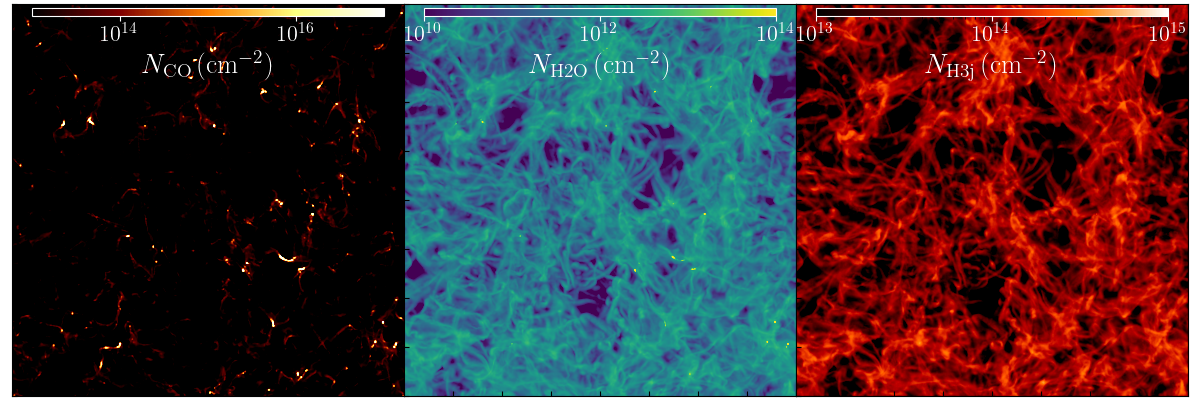}
\caption{Examples of column density maps of three important chemical species in molecular clouds, i.e. CO, H$_2$O, and H$_3^+$ respectively, from left to right. While H$_3^+$ is quite uniformly distributed in the box, H$_2$O forms in larger amounts in moderately higher density gas, and CO reaches typically observed values ($N_{\rm CO}\gtrsim 10^{17}\rm\, cm^{-2}$) only in proto-filaments (notice the huge difference, about 4 orders of magnitude, between the filaments and the background).}
\label{fig:ex_maps}
\end{figure*}

 As a final example of the unprecedented level of detail of our simulations including on-the-fly complex chemistry, we also report in Fig.~\ref{fig:ex_maps} the column density maps of relevant chemical species that are directly tracked in our simulations, i.e. CO, H$_2$O, and H$_3^+$ from left to right. CO is the mostly concentrated species, and appears in large amounts only in dense gas (proto-filaments), with the background reaching at most a 4 orders of magnitude lower abundance. Water (and especially H$_3^+$) are instead more uniformly distributed, showing mild variations across very different density conditions.

\end{appendix}

%TC:endignore

\end{document}